%% file: Double_ratio_muon_monitoring_Strategy_note_v2a.tex
\documentclass[12pt]{article}
\usepackage[usenames,dvips]{pstcol}
\usepackage{color}
\usepackage{amssymb}
\usepackage{epsfig}
\usepackage{footnote}
\usepackage{longtable}
\usepackage{verbatim}
\usepackage{array,multirow}
\usepackage{graphicx}
\usepackage{caption}
\usepackage{epstopdf}
\include{symbols}

\newrgbcolor{maroon}{.45 0.1 0.1}
\begin{document}
\pagestyle{plain}

\begin{flushright}
Version 2.02\\
\today
\end{flushright}
%
%
\begin{center}
{\Large\bf Adding Stroboscopic Muon Information For
  Reduction of Systematic Uncertainties in DUNE}\\

\vskip0.2in

Henry Frisch  
\end{center}


\begin{abstract}
 Muons have a similar latency/energy correlation from pion decay
 as do the neutrinos, and hence in each time-slice in a stroboscopic
 analysis measurements of their momentum spectra can reduce systematic
 uncertainties due to flux. There are, however, unique issues for muons: 1) during standard neutrino data-taking muon measurements in
 the forward direction must be in formidable high-flux high-radiation
 environments; 2) because of the very high incident hadron flux in the
 Absorber Hall, muons must be detected after a thick absorber,
 imposing a range cutoff at a momentum much above the minimum neutrino
 momentum of interest; 3) the muon velocity, unlike that of neutrinos,
 differs from $c$, and so the muon detected time will require
 correction for the muon flight path, requiring measurement of the
 muon momentum; 4) multiple scattering is significant for low-momentum
 muons, and so a `good geometry' is essential for precision muon flux
 measurements; and 5) developments in psec timing allow muon momenta
 in the momentum region of interest to be measured precisely by
 time-of-flight over short distances with photodetectors of a few-psec
 resolution.

 However, after trying to design arcane methods to deal with the high
 rates during routine operation, I conclude that due to the conflict
 between incident rate and absorber range cutoff, it is probably not
 possible to measure the stroboscopic muons other than in dedicated
 data-taking with the same target/horn/decay geometry, a modified
 absorber configuration with a lower range cutoff, and {\it much}
 lower proton beam intensity, itself a problem to be solved. The
 low-momentum muon spectra taken in this experiment would be
 cross-normalized to the high-intensity neutrino data through the
 currently planned muon monitors which can operate in both the low and
 high intensity geometries. Ideally the low-intensity muon momentum
 spectrum measurements would be carried out early in the LBNF program
 before the Absorber Hall becomes too hot.

While beyond the scope of uniquely muon-related issues, the note
includes a proposal for an oscillation analysis strategy that exploits
stroboscopic information for both neutrinos and muons to reduce systematic
uncertainties on the neutrino fluxes and event selection in Far and Near
detectors.

\end{abstract}

\newpage

\tableofcontents
\setcounter{tocdepth}{2}


\section{The Use of Precision Timing in Neutrino Physics}
\begin{figure}[ht]
\centering
\includegraphics[angle=0,width=1.0\textwidth]{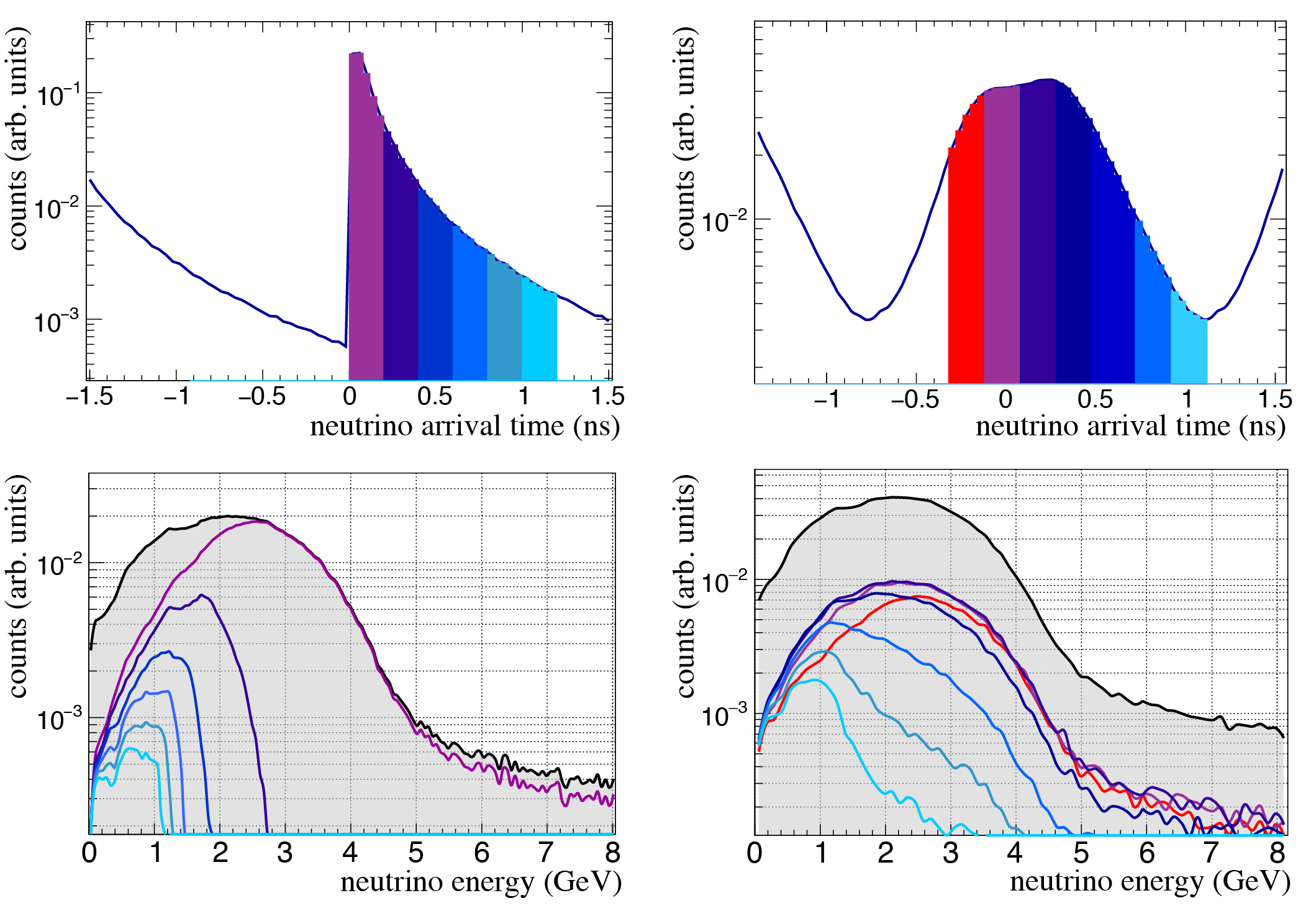}
\caption{Neutrino energy spectra in time bins relative to the proton
  bunch, for a delta-function proton bunch (left) and for a 531MHz
  bunch with 100 psec detector resolution (right). (from
  Ref~\cite{our_PRD}. Each time bin can be treated as an independent
  experiment with its own characteristic neutrino spectrum using the
  Near and Far detectors, so that five (for example) experiments are
  run simultaneously, with uncertainties on detector selection
  efficiencies in common. Here we propose to measure the momentum
  spectra of the muons associated with that time bin to reduce
  systematic uncertainties in the neutrino flux.}
\label{fig:PRD_Fig10}
\end{figure}

The further determination of parameters governing the Standard Model
neutrino sector will require much-improved control of systematic
uncertainties~\cite{deGouvea_Kelly,Ghosh,Meloni,Nosek, Ankowski}. In
response, we have proposed time-slicing the arrival of neutrinos
relative to the interaction of the proton bunch that produced them at
the target, and rebunching the Main Injector (MI) beam at a
higher-frequency such as the 10th harmonic of the current 53.1 MHz to
make shorter proton bunches~\cite{our_PRD}. This technique complements
prismatic techniques that probe the dependence on production angle of
the neutrinos~\cite{Dune_PRISM}. Figure~\ref{fig:PRD_Fig10} shows the
neutrino energy spectra in successive time bins relative to the 
nominal arrival time of the proton bunch at the target.

Here we suggest that additional constraints on the systematics may be
obtained from timing measurements of the muons produced at the target in
order to associate the muons with the time bin slices corresponding to
the muon and neutrino production\footnote{Unlike the neutrinos, the
  muons are slow, and so their detection time needs to be corrected
  for their transit time to the detector.}.

Muon momenta from pion decays will be comparable to the neutrino
momenta, and for neutrinos with momenta of a few GeV, will in general
be too soft to penetrate a thick
absorber~\cite{Groom,Clark_Kmu3}. Psec time-of-flight measurements
over lengths of 6-10 feet with small area telescopes can provide
adequate momentum resolution as well as the information needed for a
complete stroboscopic analysis that includes the muons.

However, these measurements will almost certainly require a dedicated
effort with special runs at much reduced beam intensities but with the
same target/horn/decay configuration), very-small area radiators for a
TOF telescope if used, and reduced absorber to lower the range
cutoff. Planning for precision muon monitoring in this momentum range
may affect the design and required flexibility of the absorber layout
inside the Absorber Hall. Ideally these measurements could be made
early in the LBNF program.

This proto-note is very rough and largely an aide-memoire for future
discussions. I came to the (tentative) conclusion that due to the
enormous fluxes even after a thick absorber, the high threshold set by
the absorber and the daunting constrains caused by the high-radiation
environment, that only by special runs with a much reduced proton
intensity would it be possible to make time-sliced precision measurements of
muons in the momentum range of interest. 

\section{Role of Muons in an Over-all Strategy to Reduce Systematics}
\label{muon_strategy}

The addition of neutrino timing information as described in
Ref.~\cite{our_PRD} has the capability to change the overall strategy
for reducing systematics in the long-base-line measurement. The
systematics seem~\cite{deGouvea_Kelly,Ghosh,Meloni,Nosek, Ankowski} to
depend most on 3 parameters: 1) the neutrino energy spectrum, in
particular the shape; 2) the detection efficiency for specific
signatures, such as visible energy; and 3) the K/pi ratio,
particularly for electron appearance.

A possible strategy for isolating these systematics:
\begin{enumerate}
\setlength{\itemsep}{-0.03in}

\item Treat the time bins in Figure~\ref{fig:PRD_Fig10} as 
  separate experiments, each with its own energy spectrum and
  backgrounds, so that the five experiments are run in parallel, with
  common sources of systematics but different neutrino spectra.

\item During normal data-taking, use the currently planned LBNF muon
  monitors to bin the muons in the same time bins as the neutrinos,
  making a muon sample that is the normalization for the neutrino flux
  in that time bin. These high intensity
  measurements would be cross-calibrated to the low momentum region
  directly related to the neutrino flux of interest by a separate
  dedicated experimental run with {\it much} lower beam intensity and
  with a much lower absorber range cutoff. 

\item In each experiment, bin the events by signature or characteristic
  parameter, for example electron appearance, or in visible energy, so
  that the detector efficiencies are the same\footnote{It would be
    very useful to identify sources that are not the same across the time
    bins for several signatures.}  in a given bin of the
  parameter across the five simultaneous experiments defined by the
  time bins. This leaves the spectrum and backgrounds as varying
  across the time bins, but with (approximately) the same detection
  efficiencies.

\item For example, the ratio of Far/Near in the ith time bin
  (Far/Near)$_i$ to that in the jth time bin, (Far/Near)$_j$, for a
  given visible energy, will primarily depend on the neutrino energy
  spectrum rather than on the selection and detector efficiencies. In the
  example binning of Ref.~\cite{our_PRD} there are 10 such double ratios taken
  contemporaneously for each bin of a given signature parameter. 

\end{enumerate}

\section{Some Specifically Muonic Thoughts}
Muon arrival times were already measured to a precision of 5 psec
using Cherenkov light and MCP-PMTs in 2006~\cite{Ohshima}; since then
ALD-coated MCP's with higher gain and longer life have been developed
for improved time resolution. Muons will also have an energy-time
correlation like that of neutrinos, but muons differ in that unlike
neutrinos the arrival time depends on the time the lepton travels as
well as that of the parent hadron. However, unlike neutrinos, muon
momentum can be measured locally by time-of-flight; a 630 MeV muon
($\gamma=6$) loses 14 psec per foot of travel, allowing compact
time-of-flight telescopes. Not every Main Injector spill needs to be
rebunched at 531 MHz; one in $N$ can be left at 53.1 MHz, providing
access to longer travel times for muon monitors and eliminating the
pile-up due to the higher RF frequency (`mixed-mode running). The
prescale factor $N$ need not be constant.

\subsection{The Planned LBNF Facility}
\begin{figure}[ht]
\centering
\includegraphics[angle=0,width=1.0\textwidth]{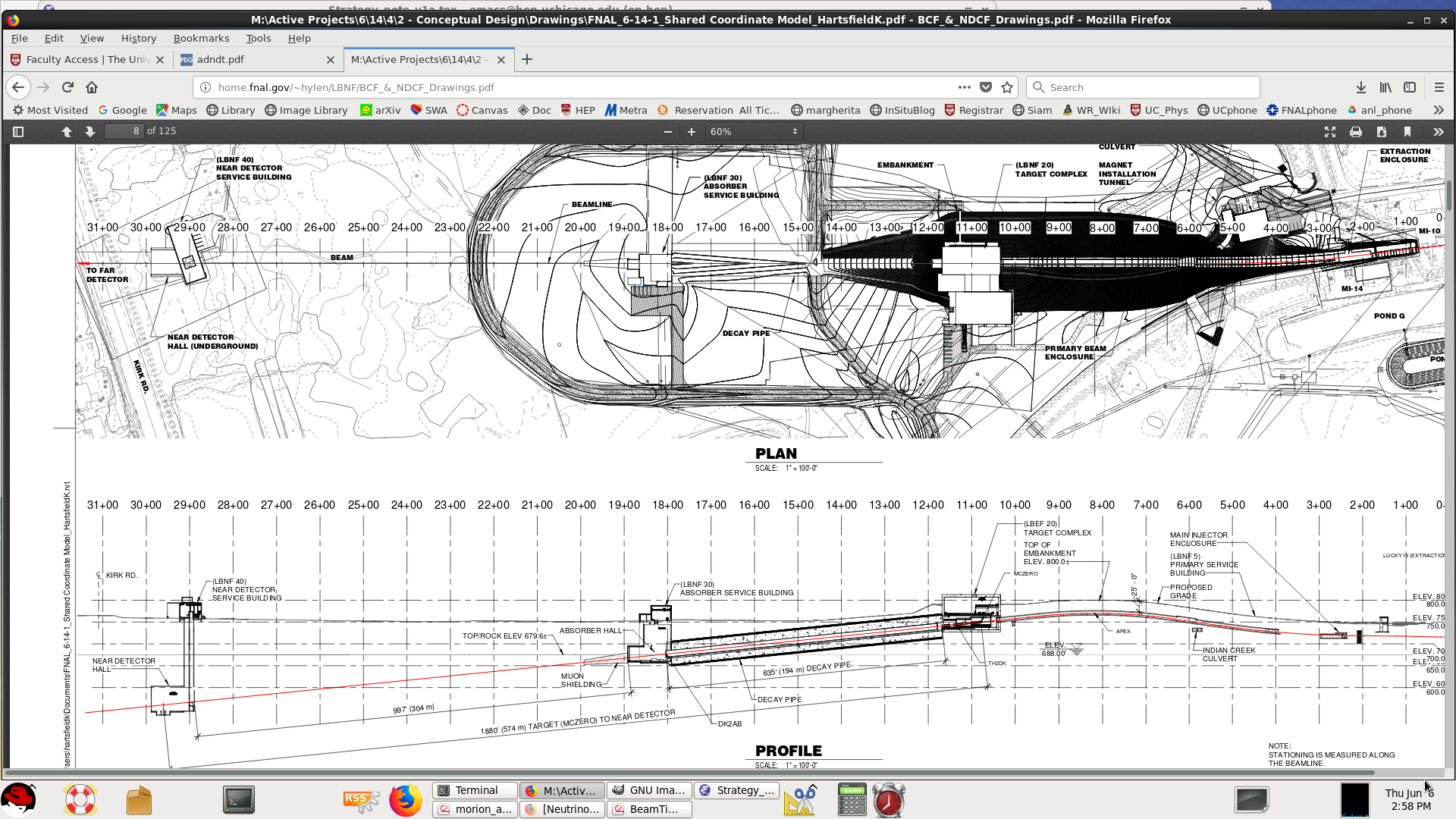}
\caption{Plan and Elevation of the LBNF target and decay region. From
  Jim Hylen's web page~\cite{Hylen}: LBNF Final Configuration, Drawing  6-14-1
  CDF-7 (page 8) }
\label{fig:Hylen_LBNF}
\end{figure}

Muon measurement is very local, and monitoring will depend 
on the details of the existing construction, radiation levels, and
availability of access.  Figure~\ref{fig:Hylen_LBNF} shows a plan view of the
facility~\cite{Hylen}. Time-of-flight at momenta of interest, however,
will require dedicated running at much lower intensities, if possible
at all.

A typical simple muon system consists of a small area (typically a few
mm-square) multi-counter telescope, with Cherenkov light read out by
PMT's or MCP-PMT's. The active element would be fused silica, for
example; the light would be read out locally or be transported to a
lower radiation environment by mirrors or phase-stabilized fibers in the case
of high radiation. Time of arrival can be measured with the same clock
distribution systems as for the neutrinos, for example White
Rabbit~\cite{white_rabbit, DUNE_timing}.

\subsection{Optical TOF Muon Momentum Measurement}
The design of a muon detection system will depend on the expected rate,
momentum range, latency (flight time relative to proton bunch), angle
to target, and background/radiation levels. The goal of accessing
muons in the kinematic ranges corresponding to the neutrinos of
interest is a really difficult one, as in the forward direction the
fluxes are enormous and require absorber that sets a momentum
threshold at much higher momentum than 1-2 GeV. New detector techniques,
including using optical time-of-flight with very small optical
elements, may allow first steps towards new monitoring designs that
provide momentum information as well as flux, and reach to lower
thresholds than before. However it seems almost certain that
measurements of the flux at momenta corresponding to that
of the neutrinos of interest will require dedicated running at much
lower intensity and lower range cutoffs, i.e. a dedicated
experiment. This need may impact the design of the placement and
flexibility of the absorber in the Absorber Hall to allow this
mode. Ideally this measurement could be carried out early in the LBNF
program before the Absorber Hall becomes too hot.

\subsection{The E100 90-Degree Monitor and the Shrinking Target}

The E100 Experiment in Proton East (PE) was a single-arm $\sim$100m-long
spectrometer designed to explore the hither-to-unexplored region of
high-$P_T$~\cite{E100}.  The spectrometer viewed targets impinged on
by the primary proton beam, with intensities up to $2E13$/spill.  Much of
the high-intensity running was done with a scintered W target.

Normalization of the measured particle production was done with a
simple ``90-degree muon monitor'' proposed and designed by Jim
Cronin. The monitor consisted of a small simple 3-counter scintillator/PMT
telescope sealed in a copper pipe and inserted into a bored/lined hole
in the ground directly above the PE target inside its heavily
shielded target box, nothing fancy. The steel shielding of the target
box and the overburden at Proton East provided a range cutoff for what
presumably were soft muons to begin with.

We noticed that the particle production rate fell with integrated
luminosity; however the rate in the 90-degree monitor fell in synch,
with the ratio deviating from being constant to at most a few percent
(this is from memory), i.e. the high-$P_T$ measurements tracked the monitor
precisely. We believed that the monitor was seeing muons
from the target, although any details of the momentum spectrum,
scattering along paths, or parent source were completely
unknown. Stability and proportionality of whatever it was, however,
was excellent.

When the run with metal targets was over, we opened up the target
box. Under the W target, which was only a few inches long if that,
there was a small conical pile of yellow dust, and the tail end of the
target had been ablated away. The 90-degree monitor, however,
tracked the shrinking target.

For a longer target such as the LBNF carbon target, multiple 90-degree
monitors with geometrically well-defined angular acceptances viewing
different sections of the target could provide similar proportionality
and stability. More-over, the new capability of measuring TOF over
short distances using fast timing detectors adds measurement of
the momentum spectrum to the monitoring.

\section{Ideas On a Linearized Oscillation Stroboscopic Analysis}
\label{analysis}

\subsection{Following Up the Stroboscopic Higher-Frequency RF Proposal
  for Physics }

The stroboscopic proposal of Ref~\cite{our_PRD} focused on the
accelerator physics of rebunching the 53 MHz of the Main Injector on
the 10th harmonic and the resulting neutrino energy spectra from a
time-sliced event selection. Missing was any estimate of 
the effect on the limiting systematic uncertainties on the neutrino
oscillation parameters. 

While beyond the scope of uniquely muon-related issues, this section
presents the bones of a proposal for an oscillation analysis that
exploits stroboscopic information for both neutrinos and muons. The
addition of fast timing at the Near and Far detectors relative to the
timing of a narrow proton bunch on target should reduce systematic
uncertainties on the neutrino fluxes and detection parameters.

\subsection{Exploiting Ratios of Time Bins in Signature Parameter
  Bins} The following strategy, expanded on the presentation in
Section~\ref{muon_strategy}, is intended to exploit the sculpted energy
spectra in the different time bins illustrated in
Figure~\ref{fig:PRD_Fig10}:

\begin{enumerate}
\setlength{\itemsep}{-0.03in}
\item Treat the time bins in Figure~\ref{fig:PRD_Fig10} as separate
  experiments, each with its own energy spectrum and backgrounds, so
  that the five time bins correspond to five oscillation experiments
  run simultaneously, with many common sources of systematics but with
  different neutrino spectra.

\item Using the muon monitors, bin the muons in the same time bins as
  the neutrinos, and use the low-intensity/high-intensity
  cross-calibration to make a muon normalization flux sample that
  corresponds to each experiment.

\item In each experiment, bin the events by signature or
  characteristic parameter, for example electron appearance or visible
  energy, so that the detector efficiencies are the same to
  first-order in a given selection bin across the five experiments
  defined by the time bins. This leaves the spectrum and backgrounds
  as varying across the time bins, but with (approximately) the same
  detection efficiencies.

\item For example, the ratio of Far/Near in the ith time bin
  (Far/Near)$_i$ to that in the jth time bin, (Far/Near)$_j$, for a
  given visible energy, will primarily depend on the neutrino energy
  spectrum rather than on the selection and detector efficiencies. In the
  example binning of Ref.~\cite{our_PRD} there are 10 such double ratios taken
  contemporaneously for a given signature.
  
\item While the example above is for one bin in a simple selection
  parameter, visible energy, the above double ratios can be calculated
  for each bin in each of the selection criteria, forming a 
matrix\footnote{Each ratio will usually have more than two indices,
    i.e. is a tensor. But...}  of
  double-ratios with the $ij$th time bin being one dimension and the
  bin of the selection parameter being the other.
  
\item Lastly, find the best-fit physics parameters and systematic uncertainties
  by comparing the complete set of measured double-ratios to
  the corresponding set of simulated predictions as a function of: 1) the
  physics parameters; 2) the flux parameterizations; and 3) the
  detector/selection efficiencies, starting with a simple
  minimization, and (inevitably) something more sophisticated and opaque.

\end{enumerate}

\section{Summary}
This draft has some ideas on strategies to reduce systematics to the
percent level by incorporating timing into the muon monitor
information.  A first step toward a solid proposal would be to analyze
the muon information in the DUNE flux simulations in a stroboscopic
framework, and to summarize expected muon rates and spectra from
existing LBNF studies.

The strategy to include muons includes:
\begin{enumerate}
\setlength{\itemsep}{-0.03in}

\item Treating each of the five 100-psec time bins (to take the
  example in the PRD draft~\cite{our_PRD}) as a separate
  contemporaneous oscillation analysis with its own neutrino spectrum
  and muon normalization. 

\item Muon flux measurements that include a dedicated running period
  early in the LBNF program to measure muon momentum spectra and
  arrival times in the momentum region that corresponds to that of the
  neutrinos of interest, using small aperture range-TOF telescopes of
  a simple design running at much lower intensity but with the same
  beamline/target/horn/decay settings and with an absorber with a
  range cutoff of at most a few GeV.

\item Measurements of muon production in
  each time bin, including angular distributions, momentum spectra, and
  (possibly) sign information to substantially constrain the
  flux/cross-section systematics.

\item Multiple 90-degree monitors with geometrically well-defined
  angular acceptances viewing different sections of the target can
  provide long-term target monitoring with proportionality and stability. Psec
  time-of-flight can measure the momentum spectrum for slow muons.

\end{enumerate}

Lastly, Section~\ref{analysis} of the note presents an outline of a ratio-based
analysis strategy that relies on stroboscopic information for both
neutrinos and muons to reduce systematic uncertainties on the neutrino
fluxes and detection parameters.

\section{Acknowledgements}
I thank Sacha Kopp, Jim Hylen, and Zarko Pavlovic for taking the time
to deal with my ignorance, and Jim and Zarko for drawings. I thank
Evan Angelico and Andrey Elagin for helpful comments on an earlier
draft, and Ed
Blucher for a comment on monitoring target stability. All
mistakes and stupidities are my own.

\section{Appendix A: Site Physical Layout Drawings}
\begin{figure}[ht]
\centering
\includegraphics[angle=0,width=0.8\textwidth]{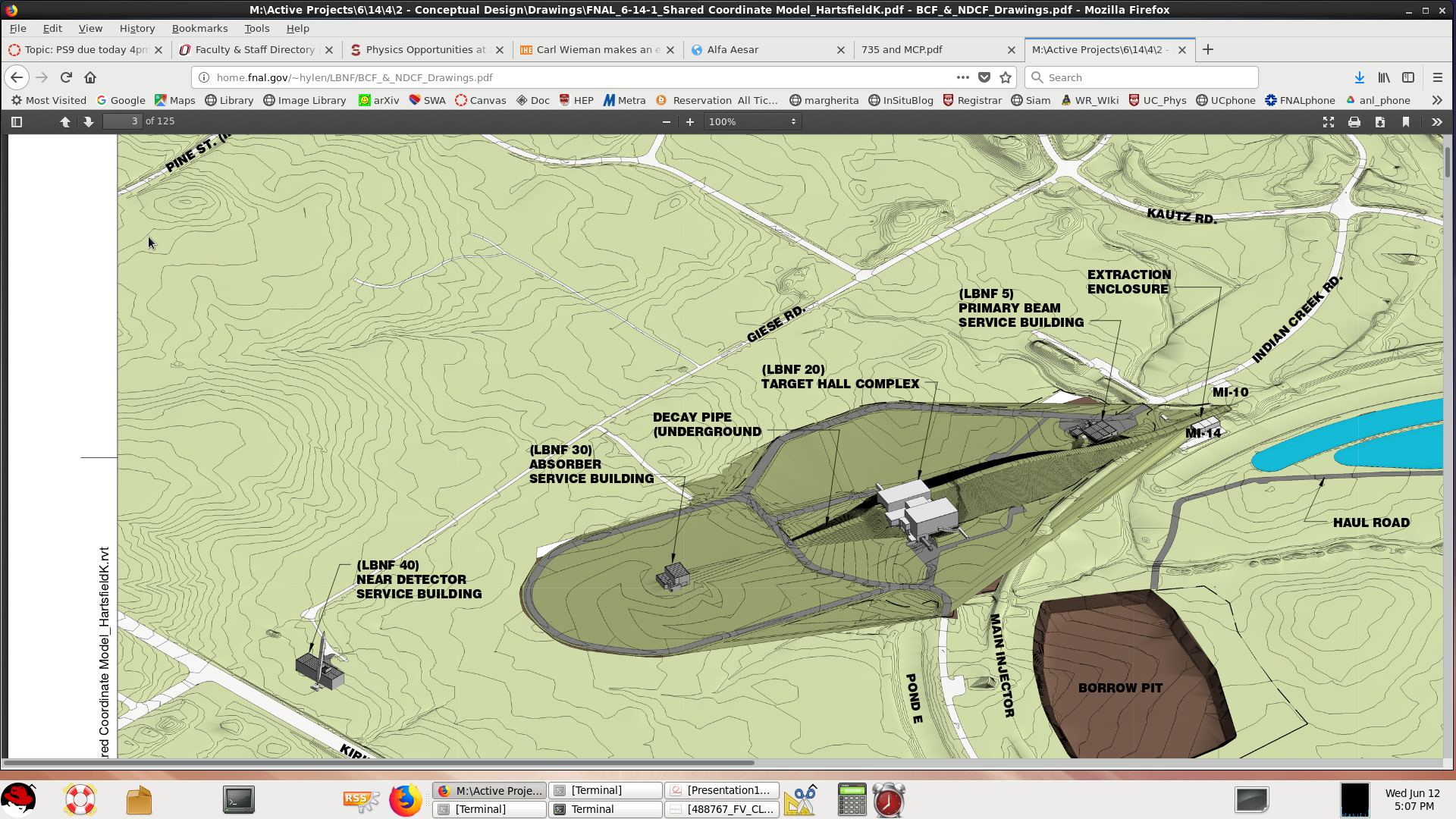}
\caption{The layout of the Fermilab site encompassing the Primary Beam
Service Building (LBNF 5), the Target Hall Complex (LBNF20), the Decay
Pipe, and the Absorber Service Building (LBNF30). See Ref~\cite{Hylen}.}
\label{fig:hylen_site_layout_color}
\end{figure}

The area comprising the target, decay pipe, and the enclosure at the
end of the decay pipe, the Absorber Hall, is shown in 
Figure~\ref{fig:hylen_site_layout_color}~\cite{Hylen}.

\begin{figure}[ht]
\centering
\includegraphics[angle=0,width=0.8\textwidth]{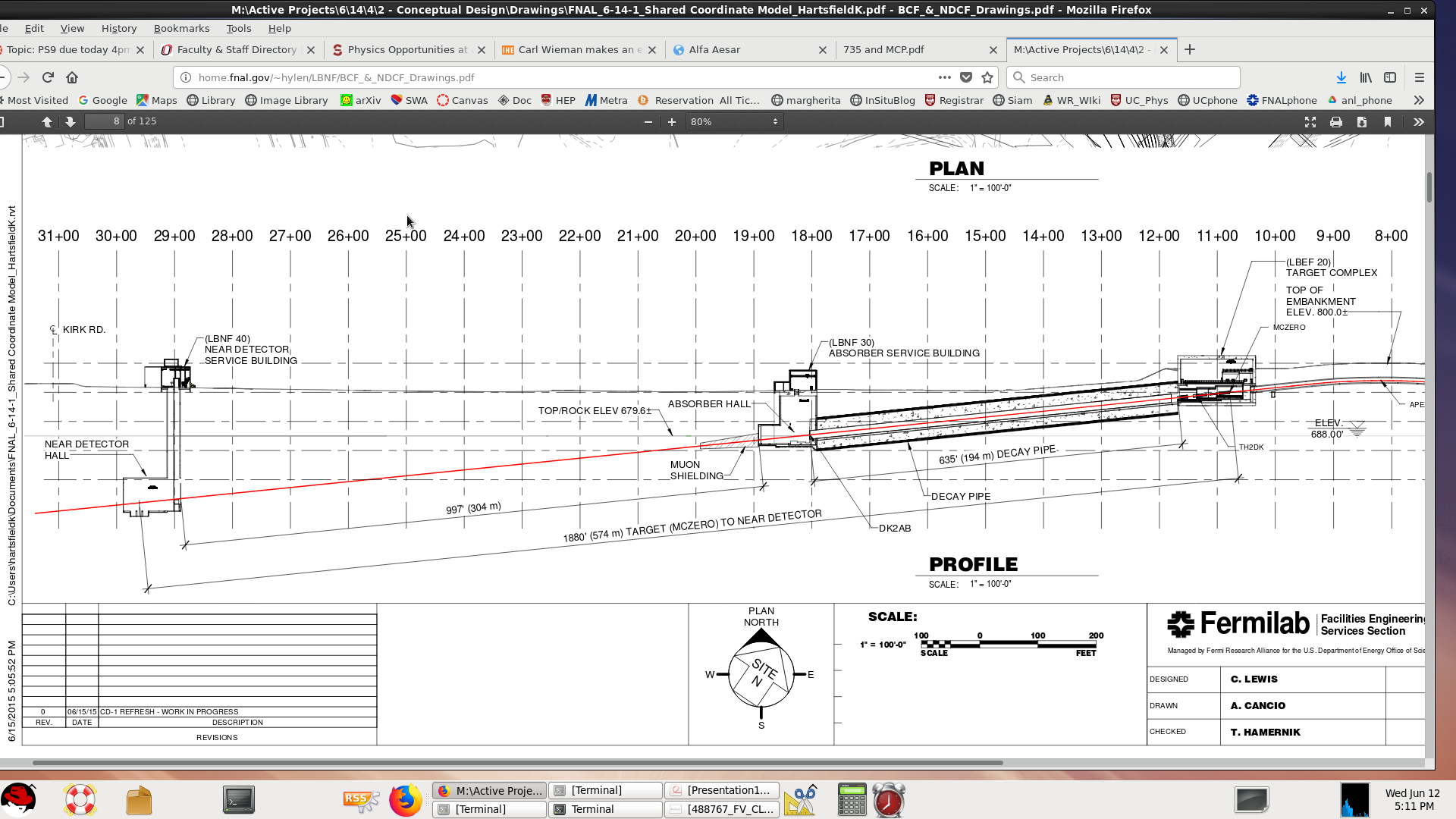}
\caption{A profile of the LBNF beamline showing the elevations of the
  highly sloped beamline, service  buildings (Target complex, Absorber
  Service Building, Near Detector Hall), and muon shielding. See Ref~\cite{Hylen}.}
\label{fig:hylen_system_profile}
\end{figure}

\clearpage
\subsubsection{Target Hall}
\begin{figure}[h]
\centering
\includegraphics[angle=0,width=0.85\textwidth]{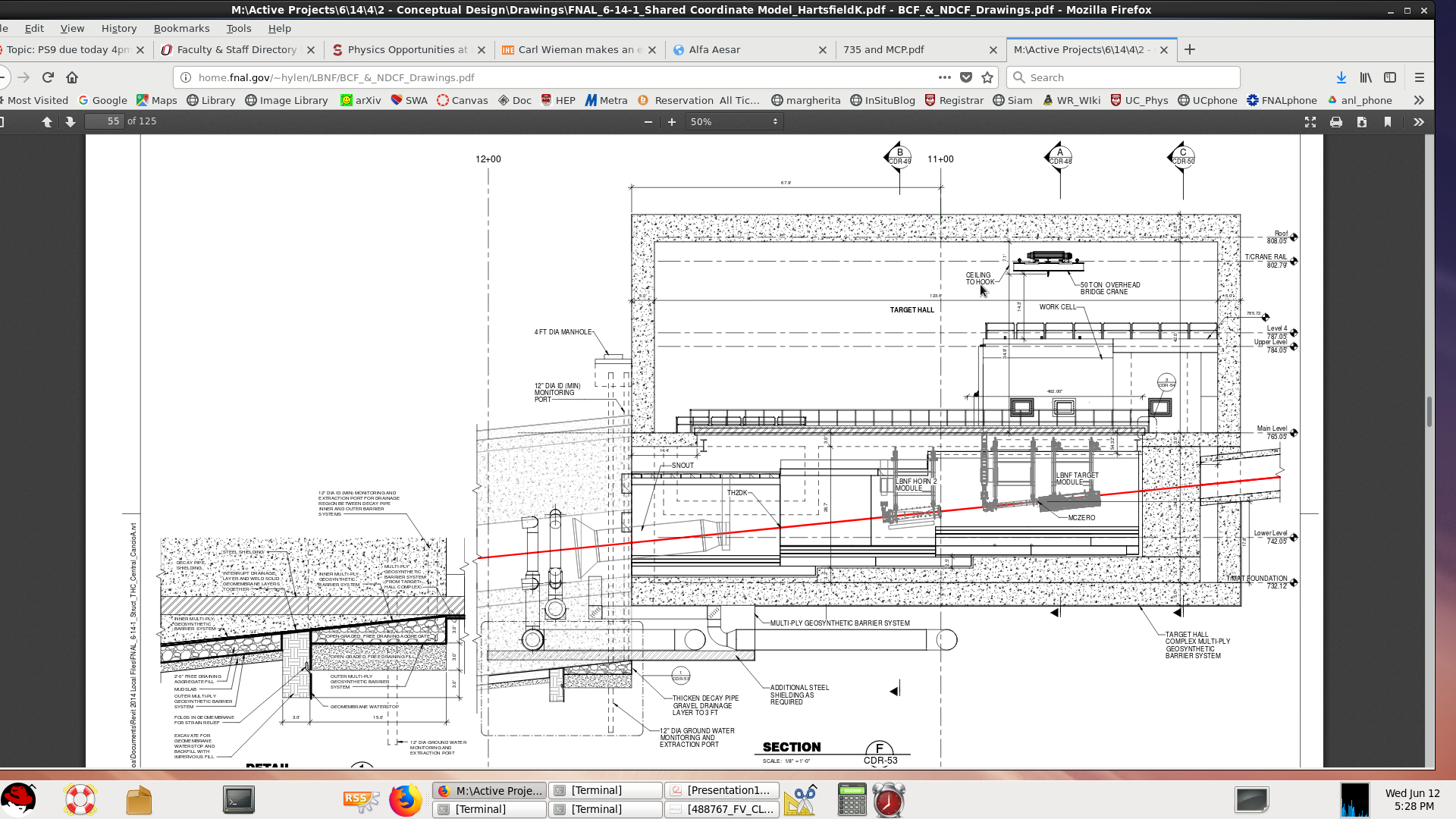}
\caption{Profile of the Target Hall.}
\label{fig:target_hall_profile}
\end{figure}

\begin{figure}[hb]
\centering
\includegraphics[angle=0,width=0.85\textwidth]{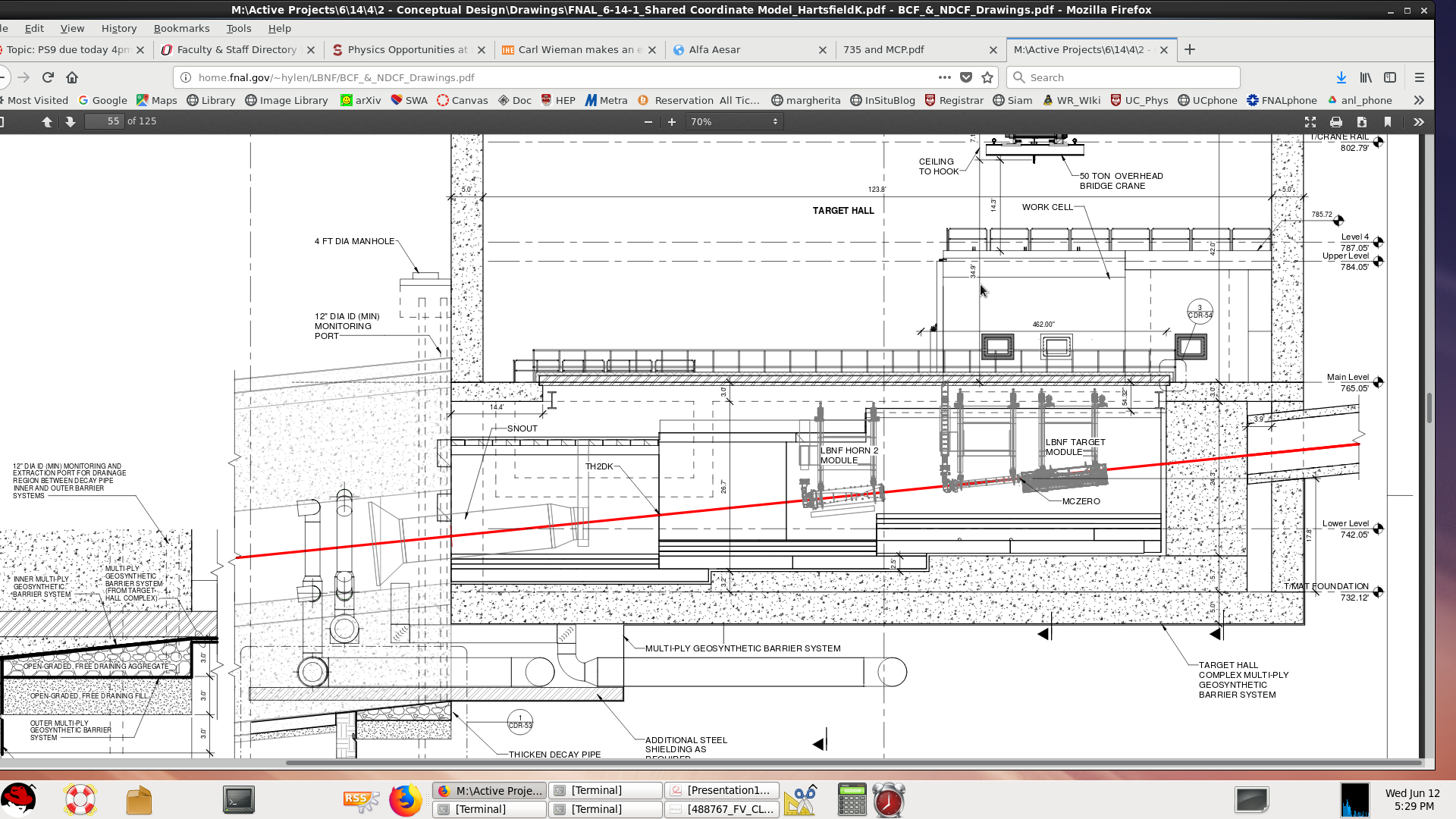}
\caption{Enlarged profile of the Target Hall.}
\label{fig:target_hall_profile_blowup}
\end{figure}

We are interested in neutrinos with momenta of order 1-2 GeV, with some
emphasis on the 2nd maximum at 800 MeV. The associated muons are
consequently also at low momentum,.
Figures~\ref{fig:target_hall_profile} and \ref{fig:target_hall_profile_blowup}
show profiles of the Target Hall. 

\clearpage
\subsubsection{Absorber Hall}

\begin{figure}[ht]
\centering
\includegraphics[angle=0,width=1.0\textwidth]{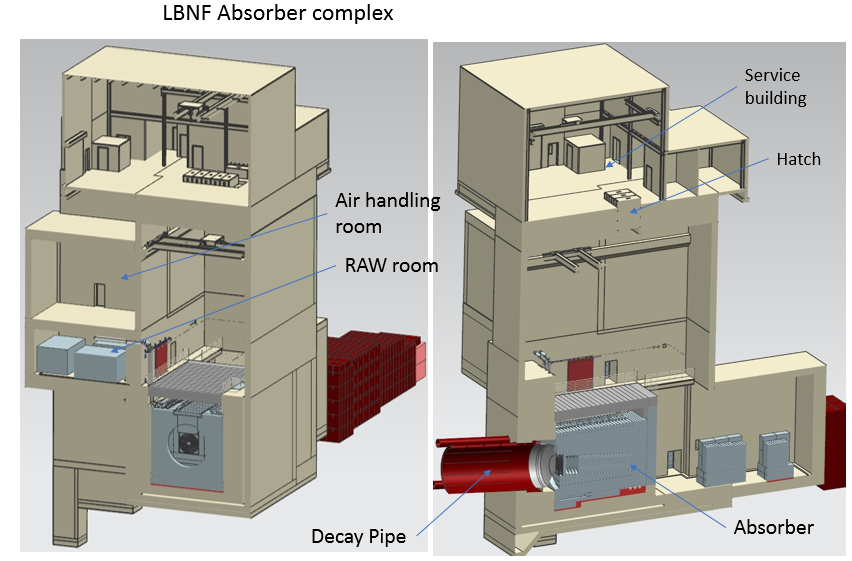}
\caption{The Absorber Hall}
\label{fig:absorber_hall_3d}
\end{figure}

\begin{figure}[ht]
\centering
\includegraphics[angle=0,width=1.0\textwidth]{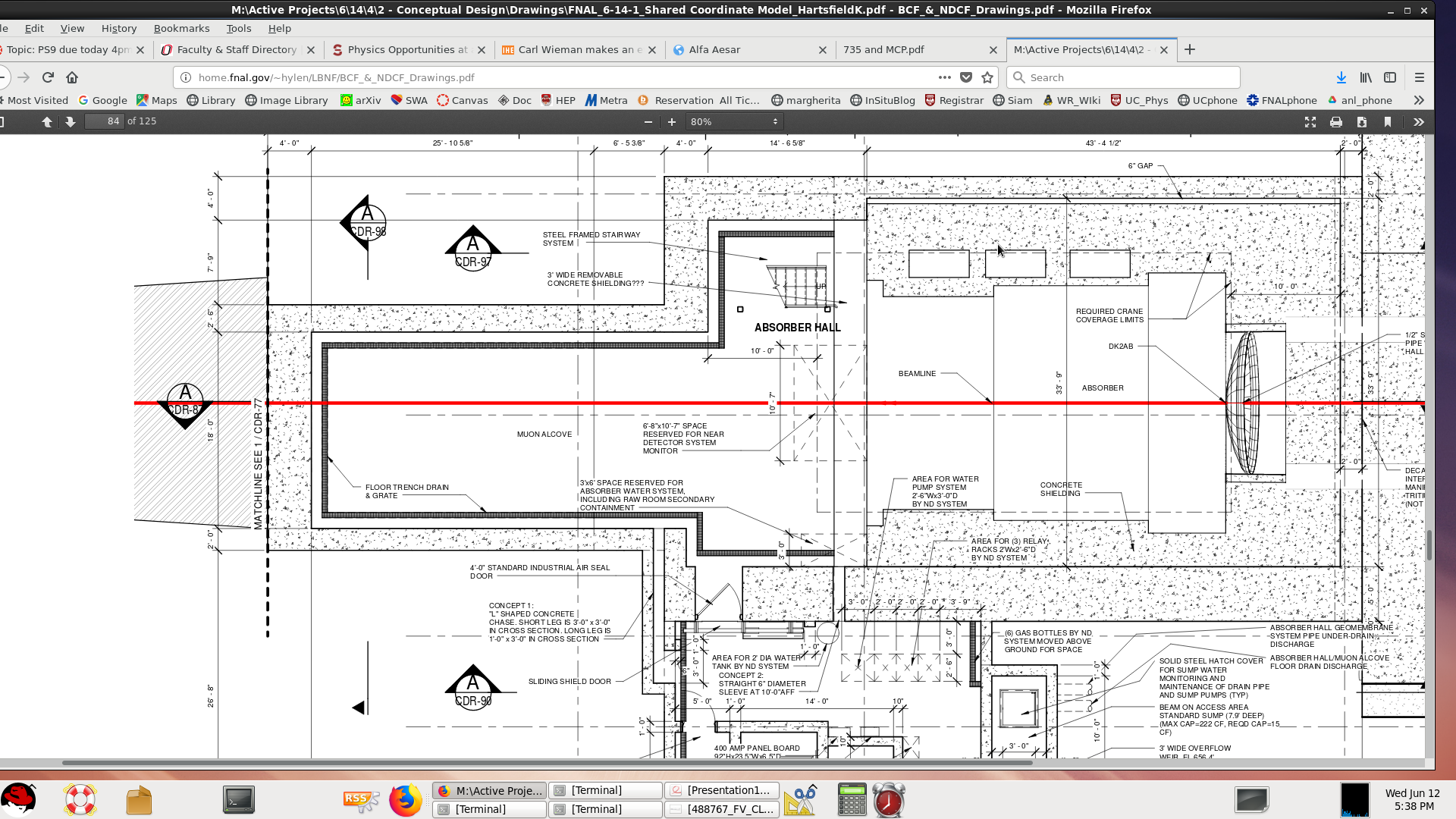}
\hfil
\includegraphics[angle=0,width=1.0\textwidth]{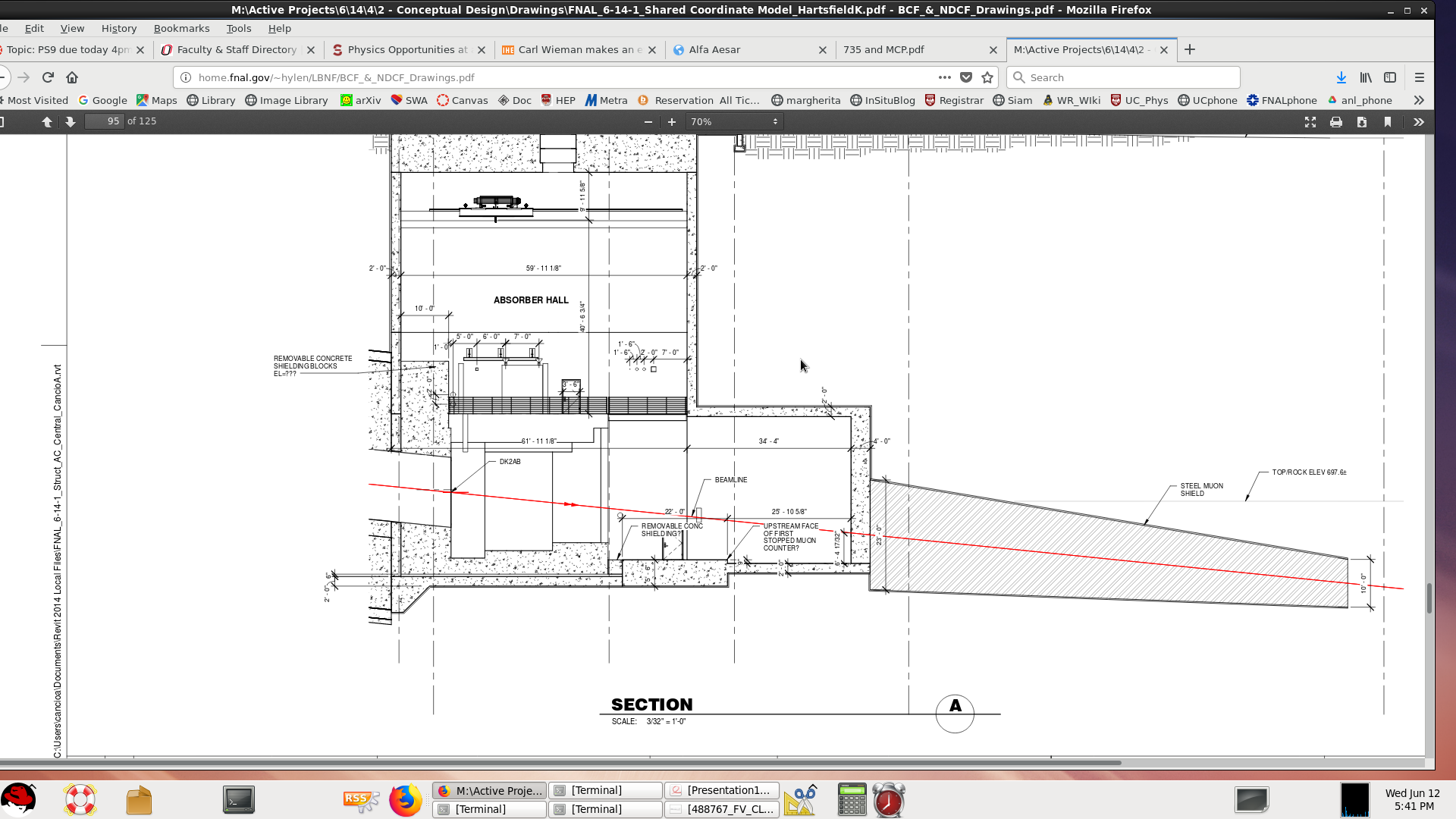}
\caption{Sections of the Absorber Hall}
\label{fig:absorber_hall_plan_stopped_muon_counter}
\end{figure}

\begin{figure}[ht]
\centering
\includegraphics[angle=0,width=1.0\textwidth]{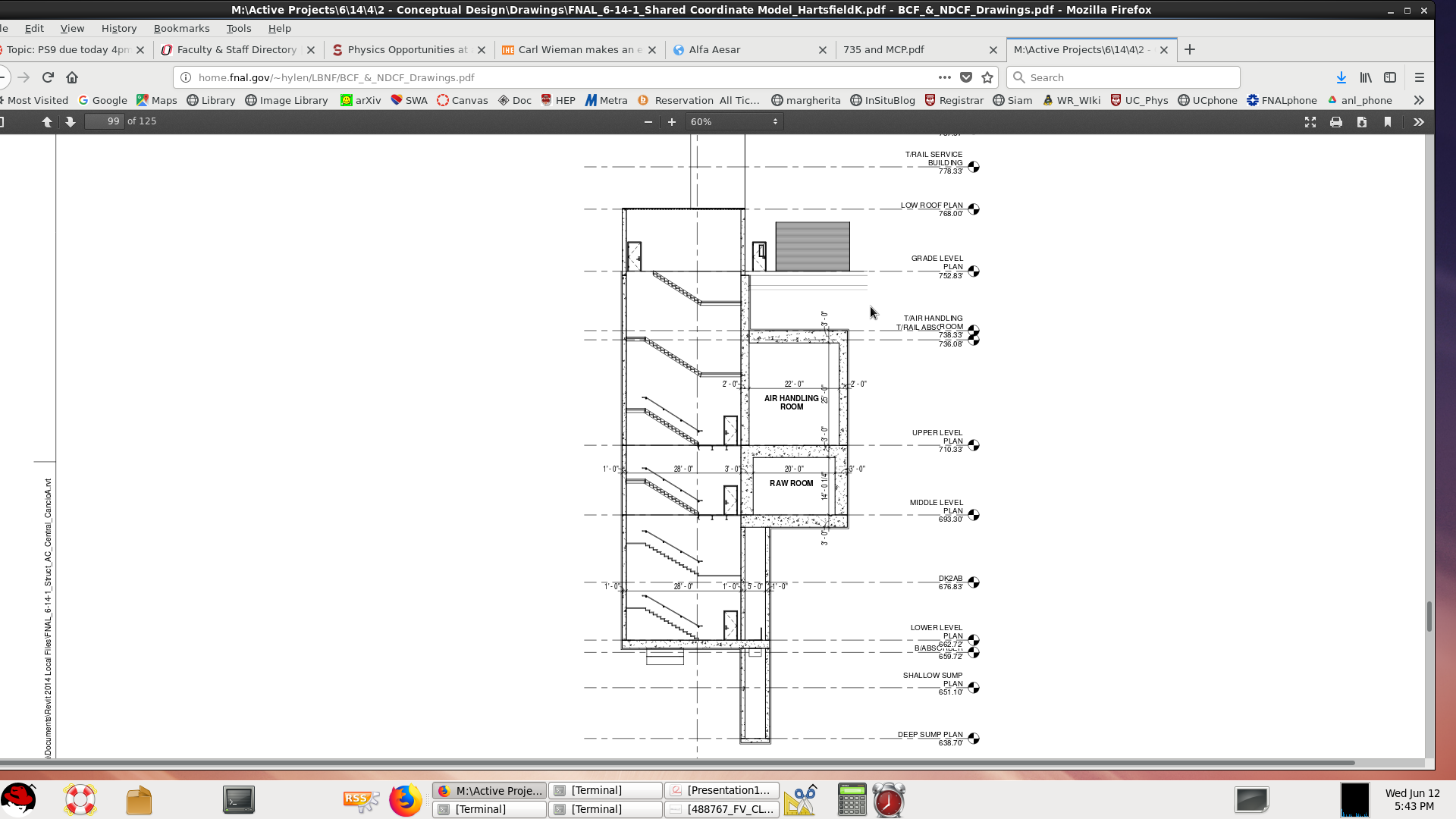}
\caption{A section view of the multi-level structure adjacent to the Absorber Hall}
\label{fig:absorber_hall_multilevel}
\end{figure}

Figures~\ref{fig:absorber_hall_3d}
--\ref{fig:absorber_hall_multilevel} show the Absorber Hall. 
Low-intensity muon flux measurements may require the ability to put
detectors in front of some fraction of the absorber and possibly 
more space before the absorber.

\clearpage

\end{document}

%% file: symbols.tex
\def \epage {\enlargethispage*{2.0in}}
\def \epage2 {\enlargethispage*{2.0in}}
\def \epage5 {\enlargethispage*{5.0in}}
\def \bc {\begin{center}}
\def \ec {\end{center}}

\def \Del \nabla
\def \Delsq \nabla^2
\def \del \partial
\def \delsq \partial^2
\def \hbar {\not h}
\def \ppm $\pm $
\def \D0 {D\O}

\def\GeVc2{GeV/{c^2}}

\def\nb-1{nb^{-1}}
\def\pb-1{pb^{-1}}
\def\fb-1{fb^{-1}}

\def\Z{${\em Z }$}

\def\goes{\rightarrow}

\def\Z0{{ Z^0}}

\def\95cl{95 \%~C.L.}
\def\95CL{95 \%~C.L.}
\def\r#1 {$^{#1}$}

\def\M4j{M_{4j}}
\def\M12{M_{12}}
\def\M34{M_{34}}